\documentclass[12pt]{article}
\usepackage{epsfig,amsfonts,amssymb}
\usepackage{hyperref}
\usepackage{cite}
\input epsf.sty
\usepackage{cite}

\def\ZZZ{{\hbox{ Z\kern-1.6mm Z}}}
\def\RRR{{\hbox{ R\kern-2.4mm R}}}
\def\CCC{{\hbox{ C\kern-2.0mm C}}}
\def\zzz{{\hbox{z\kern-1mm z}}}

\newcommand{\qeq}{{\hbox{=\kern-2.3mm ? \kern.5mm }}}
\renewcommand{\qeq}{=}

\newcommand{\eps}{\epsilon}

\newcommand{\OO}{{\cal O}}

\newcommand{\LL}{{\cal L}}

\newcommand{\wt}{\widetilde}
\newcommand{\wh}{\widehat}

\newcommand{\NN}{{\cal N}}

\newcommand{\crh}{\check\rho}
\newcommand{\cs}{\check\sigma}
\newcommand{\cv}{\check v}

\newcommand{\be}{\begin{equation}}
\newcommand{\ee}{\end{equation}}
\newcommand{\ben}{\begin{eqnarray}\displaystyle}
\newcommand{\een}{\end{eqnarray}}

\newcommand{\refb}[1]{(\ref{#1})}
\newcommand{\p}{\partial}
\newcommand{\sectiono}[1]{\section{#1}\setcounter{equation}{0}}

\def\one{{\hbox{ 1\kern-.8mm l}}}
\def\zero{{\hbox{ 0\kern-1.5mm 0}}}

\begin{document}

\baselineskip 24pt

\begin{center}
{\Large \bf  Arithmetic of Quantum Entropy Function}

\end{center}

\vskip .6cm
\medskip

\vspace*{4.0ex}

\baselineskip=18pt

\centerline{\large \rm   Ashoke Sen }

\vspace*{4.0ex}

\centerline{\large \it Harish-Chandra Research Institute}

\centerline{\large \it  Chhatnag Road, Jhusi,
Allahabad 211019, INDIA}

\vspace*{1.0ex}
\centerline{E-mail:  sen@mri.ernet.in, ashokesen1999@gmail.com}

\vspace*{5.0ex}

\centerline{\bf Abstract} \bigskip

Quantum entropy function is a 
proposal for computing the entropy
associated with the horizon of a black hole in the extremal limit, 
and is related via AdS/CFT correspondence
to the dimension of the Hilbert space in a dual quantum mechanics. 
We show that in $\NN=4$ supersymmetric string theories,
quantum entropy
function formalism naturally explains the origin 
of the subtle differences between 
the microscopic degeneracies of 
quarter BPS dyons
carrying
different torsion, \i.e.\ different
arithmetical properties. 
These arise from additional
saddle points in the path integral --
whose  existence  depends
on the arithmetical properties of the black hole charges --
constructed as 
freely
acting orbifolds of the
original $AdS_2\times S^2$ near horizon geometry.
During this analysis we demonstrate that the 
quantum entropy function
is insensitive to the details of the infrared cutoff used in
the computation, and the details of the boundary terms added to
the action. We also
discuss the role of the asymptotic symmetries of $AdS_2$ in 
carrying out the path integral in the definition of quantum entropy
function.
Finally we show that even though quantum entropy
function is expected to
compute the absolute degeneracy in a given charge 
and angular momentum sector,
it can also be used to compute the index. This
can then
be compared with the microscopic computation of the index.

\vfill \eject

\baselineskip=18pt

\tableofcontents

\sectiono{Introduction}

Extremal black holes\footnote{Throughout this paper we shall
use the word extremal black hole to denote the extremal limit of a
non-extremal black hole as reviewed in 
\cite{0805.0095,0809.3304}.}
provide us with a very useful laboratory for
understanding the quantum aspects of black hole 
physics in string theory\cite{9601029}. 
In particular one expects that for supersymmetric extremal black
holes one should be able to make a precise comparison between the
macroscopic and the microscopic entropies.
Let $d_{micro}(\vec q)$ denote the degeneracy of 
BPS microstates carrying total charge $\vec q$ in any
string theory. Then on general grounds one expects the
following relation between $d_{micro}(\vec q)$ and the
macroscopic quantities associated with 
the black hole:
\be \label{e1}
d_{micro}(\vec q)=d_{macro}(\vec q)\, ,
\ee
\be \label{e1x}
d_{macro}(\vec q) = \sum_n\,
\sum_{\{\vec q_i\}, \vec q_{hair}\atop \sum_{i=1}^n 
\vec q_i+ \vec q_{hair}=\vec q} 
\, \left\{\prod_{i=1}^n \, d_{hor}(\vec q_i)\right\}  \,
d_{hair}(\vec q_{hair}; \{\vec q_i\})\, .
\ee
The $n$-th term on the right hand side of \refb{e1x} represents
the contribution to the degeneracy from an $n$-centered black hole
configuration.
$d_{hor}(\vec q_i)$ is the degeneracy 
associated with the horizon of
a single centered black hole (or any other black object)
carrying charge $\vec q_i$, and 
$d_{hair}(\vec q_{hair};\{\vec q_i\})$ is the degeneracy of
the hair\cite{0901.0359}, 
carrying total charge $\vec q_{hair}$, 
of an $n$-centered black hole whose horizons 
carry charges $\vec q_1,\vec q_2, \cdots, \vec q_n$.
In order to make \refb{e1x} concrete 
we shall always work in a fixed duality frame. Once such a frame is
fixed, the notion of a classical solution has a well defined meaning: it
is a solution to the classical equations of motion without any 
external source
term. 
A black hole in this
duality frame will refer to
a solution to the {\it classical equations of motion} with non-singular
near horizon geometry. 
On other other hand the `hair'  of a black hole
will refer to
normalizable fluctuations of the black
hole solution with support outside the black hole 
horizon. 
For BPS black holes $d_{hair}$ can be computed by 
quantizing these normalizable fluctuations and identifying the subset
of
states which satisfy the BPS condition.
Since the space of normalizable fluctuations 
of a (multi-centered) black hole
could depend on the charges carried by various centers,
$d_{hair}$ can
depend 
on the charge $\vec q_{hair}$ of the hair
as well as the charges $\{\vec q_i\}$ of the horizons. On the other hand
since
an infinite throat separates
each horizon from the rest of the space-time, we expect the degeneracy
associated with the $i$-th horizon to depend only on the charge
$\vec q_i$ carried by that horizon, and not on the charges carried
by the other horizons or the hair.

Our main focus in this paper will be on $d_{hor}(\vec q)$.
Quantum entropy function (QEF) is a proposal for computing 
$d_{hor}(\vec q)$
in terms of a path integral over string fields
on the near horizon attractor 
geometry of the black hole containing a product of $AdS_2$ and
a compact space $K$. This has been described in \refb{ealt}.
We shall use \refb{ealt} to analyze the QEF
of quarter BPS black holes in heterotic string theory compactified on
$T^6$ and compare this with the microscopic prediction.
This theory has $O(6,22;\ZZZ)$ T-duality group which is
generated by $O(6,22;\RRR)$ matrices preserving the 28
dimensional Narain lattice\cite{narain}, 
and the dyons are characterized by
28 dimensional electric and magnetic charge vectors $(Q,P)$
taking values in the Narain lattice.
It has been known for sometime that the microscopic degeneracy
of the dyons, besides depending on the 
invariants $(Q^2,P^2,Q\cdot P)$
of the continuous $SO(6,22;\RRR)$ group, also depends
on
\be \label{eip1}
\ell = \gcd\{Q_i P_j - Q_j P_i\}\, ,
\ee
which is an invariant of the discrete T- and S-duality 
groups\cite{0702150,0712.0043,0801.0149} encoding arithmetical
information about the charges.  We shall refer to
this integer as torsion.
Using
an appropriate S-duality transformation a dyon of torsion
$\ell$ can be brought to the form $(\ell\, Q_0, P_0)$ where 
$Q_0$, $P_0$ are primitive lattice vectors satisfying 
$\gcd\{Q_{0i} P_{0j} - Q_{0j} P_{0i}\}
=1$\cite{0801.0149}. The degeneracy
of such dyons is given 
by\cite{0802.0544,0802.1556,0803.2692}
\be \label{ech2int}
(-1)^{Q\cdot P+1}
\sum_{s|\ell} s\, f(Q^2/s^2, P^2, Q\cdot P/s)\, ,
\ee
where $(-1)^{Q\cdot P+1}
f(Q^2, P^2, Q\cdot P)$ denotes the degeneracy of a dyon
of charge $(Q,P)$ with $\gcd\{Q_iP_j-Q_jP_i\}=1$. 
Our goal will be to understand this formula from a macroscopic
viewpoint, \i.e.\ by using the quantum entropy 
function.\footnote{For earlier discussion on the relation between
attractor geometry and arithmetic
see \cite{9807056,9807087,0401049}.}

The function $f$ is given by the triple Fourier transform of the
inverse of the Igusa cusp 
form\cite{9607026,0412287,0505094,0506249,0605210}. 
In principle QEF should provide a complete macroscopic
derivation of the
function $f$ as well as the structure of eq.\refb{ech2int}.
However our goal in this paper will be modest; instead of
providing a detailed derivation of the function $f$ from QEF, 
we shall simply try to identify the 
origin of different terms in the sum in \refb{ech2int}.
We shall show that there are natural candidates which reproduce these
terms, -- they reflect contribution from different saddle points with the
same asymptotic field configuration as the
near horizon attractor 
geometry of the black hole. These new saddle points, obtained as
freely acting $\ZZZ_s$ orbifolds of the original near horizon attractor
geometry, are
in one to one correspondence with the divisors of $\ell$, and
furthermore the classical contributions to QEF 
from these saddle points coincide with the leading 
asymptotic behaviour of the summands $f(Q^2/s^2, P^2, Q\cdot P/s)$
for large charges.

In this context we would like to remind the reader that the
function $f(Q^2,P^2, Q\cdot P)$ itself can be expressed as a sum over
infinite number of terms, associated with the infinite number of poles
of the Igusa cusp form, and it was argued in \cite{0810.3472} 
that these different
terms can be associated with different saddle points obtained by
taking the quotient of the original $AdS_2\times S^2$ background
by $\ZZZ_N$ orbifold groups. 
These orbifolds exist for all charges, including those with unit
torsion.
The saddle points considered here
are distinct from the ones used in \cite{0810.3472} due to inclusion
of additional shift transformations in the orbifold group action, 
and exist only for dyons
carrying non-trivial torsion. 

The analysis of QEF for dyons of
$\NN=4$ supersymmetric string theory has been described in
\S\ref{storsion}. However
we also
address several technical issues related to the
computation of QEF, a summary of which
is given below.
\begin{enumerate}
\item Since $AdS_2$ has infinite volume, path integral over string
fields in $AdS_2$ suffers from infrared divergences as in the case of
higher dimensional $AdS$ spaces\cite{9802109,9802150}. 
Thus in order to get sensible answers one first
needs to use an infrared cutoff that regularizes the volume of $AdS_2$
and at the end of the computation take the cutoff to infinity. In earlier
work\cite{0805.0095,0809.3304} 
the infrared cutoff was chosen in a way so that it preserves
an $SO(2)$ subgroup of the $SL(2,R)$ isometry of $AdS_2$. However
one can choose a more general infrared cutoff that destroys all isometries
of $AdS_2$. We show that even with such  general infrared cutoff
one gets exactly the same value of QEF. Thus QEF 
is insensitive to the details of the 
infrared cutoff. We also show that QEF is insensitive to the details
of the boundary terms which are added to the action.

\item Besides the $SL(2,R)$ isometry, $AdS_2$ has an infinite group
of asymptotic symmetries. We need to take special care 
in defining the path
integral so that integration
over these symmetry directions do not generate an infinite factor. We
discuss this in detail and give a specific prescription for removing these
infinite factors from the path integral.

\item 
Via $AdS/CFT$ 
correspondence\cite{9711200,9802109,9802150,9809027} 
one can argue that QEF counts the number of ground states of
the black hole in a given charge and angular momentum sector after
removing the contribution from the hair degrees of 
freedom\cite{0809.3304}.
Now often in the comparison between the macroscopic and the 
microscopic entropies one computes an index rather than 
the absolute
degeneracy on the microscopic side
since it is this index that is protected by supersymmetry.
In particular the `degeneracy formula' given in \refb{ech2int}
actually refers to the sixth helicity trace 
$-B_6$\cite{9708062,9708130}. This raises a question as to how
QEF can be compared with the index computed
on the microscopic side.
We show that since QEF measures the
degeneracy in fixed charge and angular momentum sector, it
can actually be used to compute an index on the
macroscopic side. This can then be compared with the
microscopic index.

\end{enumerate}

\sectiono{Quantum Entropy Function} \label{squant}

In this section we shall give a brief overview of the quantum entropy
function, -- the quantity that is supposed to compute $d_{hor}(\vec q)$
appearing in \refb{e1x}.
We begin by writing down 
the background fields describing the
$AdS_2$ near horizon geometry of an extremal 
black hole\cite{0805.0095,0809.3304}:
\be \label{et2}
ds^2
= v\left(-(r^2-1) dt^2+{dr^2\over
r^2-1}\right),  \quad F^{(i)}_{rt} = e_i, \quad \cdots
\ee
where $F^{(i)}_{\mu\nu} = \p_\mu A^{(i)}_\nu - 
\p_\nu A^{(i)}_\mu$ are the gauge field strengths, $v$ and $e_i$ are
constants and $\cdots$ denotes near horizon values of other fields.
Under euclidean continuation
\be \label{et6}
t = -i\theta\, ,
\ee
we have
\be \label{et6.5}
ds^2
= v\left((r^2-1) d\theta^2+{dr^2\over
r^2-1}\right),  \quad F^{(i)}_{r\theta} = -i \, e_i, \quad \cdots
\ee
Under a further coordinate change
\be \label{et6.6}
r = \cosh\eta\, \, ,
\ee
\refb{et6.5} 
takes the form
\be \label{et7}
{ds^2} = { v \, \left(d\eta^2 +\sinh^2\eta \, d\theta^2 \right),}
\qquad
F^{(i)}_{\eta\theta} = -i e_i \, \sinh\eta, \qquad \cdots \, .
\nonumber \ee
The metric is non-singular at the
point $\eta=0$ if we choose $\theta$ to have period $2\pi$.
Integrating the field strength we can get the form of the gauge
field:
\be \label{et8}
A_\mu^{(i)} dx^\mu = -i \, e_i \, (\cosh\eta  
{ -1})\, d\theta=  -i \, e_i \, (r  
{ -1})\, d\theta\, .
\ee
Note that the $-1$ factor inside the parenthesis is required to make the
gauge fields non-singular at $\eta=0$.  In writing \refb{et8} we have
chosen $A_\eta=0$ gauge. 

Quantum entropy function is now defined as
\be \label{ealt}
d_{hor}(\vec q) = \left\langle \exp[-
i  q_i\ointop d\theta \, A^{(i)}_\theta]
\right\rangle^{finite}_{AdS_2}\, .
\ee
Here $\ointop \, d\theta\, A^{(i)}_\theta$
denotes the integral of the $i$-th gauge field along the boundary
of $AdS_2$.  $\langle ~\rangle_{AdS_2}$ 
denotes the unnormalized path integral
over various fields, satisfying the same asymptotic behaviour
as \refb{et7}, weighted by $e^{-A}$ where $A$ is the Euclidean
action.
The path integral must be performed over all the
fields  in string theory.
The superscript `${finite}$' refers to the finite part of the
amplitude defined as follows.
If we regularize the infrared divergence by putting an
explicit upper cutoff on $r$ (or $\eta$), 
and denote by
$L$ the length of the boundary of this regulated
$AdS_2$, then for large cutoff, \i.e.\ large $L$,
the amplitude 
has the form $e^{C L + \OO(L^{-1})}\times\Delta$ 
where $C$ and $\Delta$ are $L$-independent
constants. The finite part of the amplitude
is defined to be the constant $\Delta$.
Eq.\refb{ealt}, together with \refb{e1}, \refb{e1x},
gives a precise relation
between the microscopic degeneracy and an
appropriate partition function in
the near horizon geometry of the black hole.
We  call the right hand side of \refb{ealt}, \i.e.\ the
constant $\Delta$, the `quantum entropy
function' (QEF).

In defining the path integral over $AdS_2$ we need to put boundary
conditions on various fields. Special care is needed to fix the boundary
condition on $A^{(i)}_\theta$.
In the $A^{(i)}_\eta=0$ gauge
the solution of the linearized Maxwell's equations around
this background has two independent solutions near the boundary: 
$A^{(i)}_\theta
={\rm constant}$ and $A^{(i)}_\theta \propto r$. Since the latter
is the dominant mode for large $r$,
we put boundary condition on the latter mode,
allowing the constant mode of the gauge field to be integrated over.
This is done formally by requiring
\be \label{efor1}
\lim_{r\to\infty}\, 
{\delta A_{bulk}\over \delta F^{(i)}_{r\theta} }= -i \, q_i
\ee  
where $\{q_i\}$ are fixed numbers and $A_{bulk}$ 
is the full bulk action of the
two dimensional theory. 
This corresponds to fixing the
charges carried by the black hole to be $\{q_i\}$. 
In this case under an infinitesimal 
variation of the
gauge field component $A^{(i)}_\theta$ we have
\be \label{efor2}
-\delta A_{bulk} = -\int \, dr\, d\theta \, \p_r \delta A^{(i)}_\theta \, 
{\delta A_{bulk}\over \delta F^{(i)}_{r\theta} }
= \int \, dr\, d\theta \, \delta A^{(i)}_\theta \, \p_r \, 
{\delta A_{bulk}\over \delta F^{(i)}_{r\theta} }
- \ointop \, d\theta \, \delta A^{(i)}_\theta \, 
{\delta A_{bulk}\over \delta F^{(i)}_{r\theta} }\bigg|_{boundary}\, .
\ee
Setting the first term on the right hand side of
\refb{efor2} to zero yields the equation of motion of
$A^{(i)}_\theta$. On the other hand the second term
cancels against the variation of the 
$-i q_i \ointop d\theta \, A^{(i)}_\theta$ term in the exponent in
\refb{ealt}.
The boundary conditions
on the other fields are fixed in the standard manner, {\it e.g.} in the
$g_{\eta\eta}=v$, $g_{\theta\eta}=0$ gauge we
freeze the mode of $g_{\theta\theta}$ proportional to $r^2$ to its
value given in \refb{et7} and allow the constant mode of the metric
to fluctuate. On the other hand for massless
scalar fields we require the constant
mode of the fields to vanish at asymptotic infinity.
Appropriate boundary terms must be added to the action so that the
variation of the action under an arbitrary variation of various fields,
subject to these boundary conditions, vanishes 
when equations of motion
are satisfied.

In this context note that we could have included the exponential term in
\refb{ealt} as part of the boundary term in the action and expressed
\refb{ealt} as
\be \label{ex1}
d_{hor}(\vec q) = Z^{finite}_{AdS_2}\, ,
\ee
where $Z_{AdS_2}$ is the partition function of $AdS_2$ 
computed using
the natural boundary condition that fixes the electric charge rather than
the gauge potential.
We shall continue to use the notation \refb{ealt} since it explicily 
displays the part of the boundary term that requires 
the gauge potential and cannot be expressed in
terms of gauge field strengths.
\refb{ealt} also has the advantage that it explicitly displays the
dependence on the charges $q_i$.

Since QEF can be regarded as the partition
function on an Euclidean $AdS_2$ in a fixed charge (including
angular momentum) sector, the usual rules of
$AdS/CFT$ 
correspondence\cite{9802109,9802150} tells us that
it measures
the partition function of a dual quantum mechanics living on the
boundary of $AdS_2$\cite{0809.3304}.
This dual quantum mechanics in turn
can be obtained as the infrared limit of the quantum mechanics
describing the dynamics of the black hole after removing its hair
degrees of freedom. Since from microscopic analysis one
finds that the black
hole has a gap separating the ground state from the first excited state,
only the ground states in a given charge and angular
momentum sector survive in the infrared limit, and the partition
function takes the form $d(q) e^{-E_0 L}$ where $d(q)$ is the
ground state degeneracy, $E_0$ is the ground state energy and
$L$ is the length of the boundary of $AdS_2$. The `finite part' of
this is given by $d(q)$. Thus QEF 
should count the number of ground states of
the black hole in a given charge and angular momentum sector after
removing the contribution from the hair degrees of freedom.
{}From this viewpoint the proposal that QEF
measures the degeneracy associated with the horizon is a
direct consequence of $AdS_2/CFT_1$ correspondence.
As a consistency check on this proposal,
it was shown in \cite{0809.3304} that QEF 
reduces to the exponential of the Wald 
entropy\cite{9307038,9312023,9403028,9502009}
in the classical limit.

\sectiono{Insensitivity to the Infrared Cutoff} \label{sinfra}

It was shown in \cite{0809.3304} that 
QEF reduces to
the exponential of the
Wald entropy in the classical limit. During this proof the
infrared divergence associated with the infinite volume of
$AdS_2$ was regularized by putting a cutoff at $r=r_0$. In 
this section we shall consider a more general cutoff of the
form\footnote{We have imposed the 
last condition in \refb{ea1} to fix the overall
normalization of $f$, since any change in the normalization of
$f$ can be absorbed into a redefinition of $r_0$.}
\be \label{ea1}
r\le r_0 \, f(\theta) \, , \quad f(\theta+2\pi)=f(\theta), 
\quad f(\theta)>0, \quad \int_0^{2\pi} d\theta f(\theta) = 2\pi\, ,
\ee
for any smooth function $f(\theta)$
and show that the result does not change. 
We begin by noting that in the classical limit QEF 
is given by the finite part of
\be \label{ecc1}
\exp\left(-A_{bulk} - A_{boundary} - i q_i \, 
\ointop A^{(i)}_\theta \, d\theta\right)\, ,
\ee
where $A_{bulk}$ and $A_{boundary}$ represent contributions from
the bulk and the boundary terms in the classical
action in the background
\refb{et7}.
If $\LL$ denotes the
Lagrangian density of the two dimensional theory, 
then the bulk contribution to the action
in the background \refb{et7} takes the form:
\ben \label{ea2}
A_{bulk} &=& - \int d^2 x \,  
\sqrt{\det g} \, \LL \nonumber \\
&=&    -\int_0^{2\pi} d\theta \, \int_0^{\cosh^{-1}(r_0 f(\theta))}\,
d \eta \, \sinh\eta\, v \, \LL \nonumber \\
&=& - v \, \LL \, 
\int_0^{2\pi} \left( r_0\, f(\theta) -1\right)\, d\theta  \nonumber \\
&=& - 2\pi \, v \, \LL \, (r_0-1) \, .
\een
In going from the second to the third step in \refb{ea2} we have
used the fact that due to the $SO(2,1)$ invariance of the $AdS_2$
background, $\LL$ is independent of $\eta$ and $\theta$.
In this parametrization the length  $L$  of the boundary is
given by
\be \label{ea3}
L = 
\int_0^{2\pi} \, d\theta\, \sqrt{r_0^2 f(\theta)^2 - 1}  =
2\pi\, r_0  + \OO(r_0^{-1})\, .
\ee
The contribution from the last term in the exponent
in \refb{ecc1} can also be
calculated easily using the expression for $A^{(i)}_\theta$
given in \refb{et8}. We get
\be \label{ea3.5}
i q_i \, 
\ointop A^{(i)}_\theta \, d\theta
= 2\pi\, \vec q\cdot \vec e \, (r_0-1)\, .
\ee
Finally,
the contribution from $A_{boundary}$ 
can be analyzed as follows. $A_{boundary}$ is chosen so that
under an arbitrary variation of the fields the boundary terms arising
in $\delta A_{bulk}$ get cancelled by the boundary terms arising 
from the variation of $A_{boundary}$. Now as we have discussed
in \S\ref{squant}, the boundary terms in $\delta A_{bulk}$ 
proportional to $\delta A^{(i)}_\theta$ are cancelled by the last
term in the exponent in \refb{ecc1}. Thus $A_{boundary}$ 
must be chosen so as to cancel the boundary
contribution to $\delta A_{bulk}$
from the variation of
the other fields, without giving any further term involving 
$\delta A^{(i)}_\theta$. This in particular requires that 
the dependence of $A_{boundary}$ 
on $A^{(i)}_\theta$ enters only through the field strengths
$F^{(i)}_{\theta \eta}$.
In order to analyze the contribution from these terms we 
choose new coordinates near the
boundary\footnote{Although it is not necessary, we
could choose the coordinate transformations near $\eta=0$ 
to look like
$\xi=\eta-\ln r_0-\ln 2$, $w=2\pi r_0\theta$, so that the solution
near the core is independent of the choice of $f(\theta)$.}
\be \label{ea5}
\xi 
=\eta - \ln r_0 -\ln 2 - \ln f(\theta), \qquad 
w = r_0 \int_0^\theta f(u)du - {1\over 2} r_0^{-1}\, e^{-2\xi}\,
{f'(\theta)\over f(\theta)^2} \, ,
\ee
so that we have, due to the equivalence $\theta\equiv \theta+2\pi$,
and the properties of $f(\theta)$ described in \refb{ea1},
\be \label{ea5.5}
w\equiv w+2\pi r_0\, ,
\ee
and   the boundary $\eta=\cosh^{-1}(r_0 f(\theta))$ is at 
\be \label{ea6}
\xi= \OO(r_0^{-2})\, .
\ee
In this coordinate system the background 
\refb{et7} near the boundary
takes
the form
\ben \label{ea8}
{ds^2} &=& v \, \left(d\xi^2 + e^{2\xi}\, 
dw^2 \right) + \OO(r_0^{-2}),
\nonumber \\\
F^{(i)}_{\xi w} &=& - i e_i  \, e^\xi  +
\OO(r_0^{-2})\, ,
\quad \cdots \, .
\een
We now note that 
the background \refb{ea8}
is independent of $r_0$ up to corrections of order
$r_0^{-2}$. 
Thus 
the integrand of the boundary term is $r_0$ independent up to terms
of order $r_0^{-2}$.
\refb{ea8} also has the translation symmetry under $w\to w+c$
up to corrections of order
$r_0^{-2}$ so that the $w$ integral in
$A_{boundary}$ will produce a factor of
$L=2\pi r_0$ up to corrections of order
$r_0^{-1}$. Thus the boundary term will give a 
contribution of the form
\be \label{eend1}
A_{boundary} = 2\pi r_0 \, K + \OO(r_0^{-1})\, ,
\ee
for some constant $K$.
Combining \refb{ea2}, \refb{ea3.5} and \refb{eend1}
we get
\ben \label{eend3}
\exp\left(-A_{bulk} - A_{boundary} - i q_i \, 
\ointop A^{(i)}_\theta \, d\theta\right)
&=& \exp\left [-2\pi r_0 (\vec q\cdot \vec e - v\, \LL +K)+
\OO(r_0^{-1})\right]\nonumber \\
&& \times
\exp\left[2\pi (\vec q\cdot \vec e - v\, \LL) \right]\, .
\een
Thus QEF, given by the finite part of
\refb{eend3}, takes the form
\be \label{eend4}
d_{hor}(q) = \exp\left[2\pi (\vec q\cdot \vec e - v\, \LL) \right]\, .
\ee
The right hand side of \refb{eend4} is the exponential of the
Wald entropy\cite{0506177}. 
Thus we see that the result is independent of
the function $f(\theta)$ we use to regulate the infrared divergence.
Also note that by changing the boundary action we can change
the coefficient $K$, and hence the coefficient of the term proportional
to $r_0$ in the exponent of \refb{eend3}, but the $r_0$ independent part
that contributes to QEF is not affected by
the choice of the boundary action.\footnote{We
could also try to adjust the boundary terms, \i.e.\ the constant
$K$,  so that the term in the exponent of \refb{eend3} linear
in $r_0$ vanishes, and we are left with only the finite part of
the amplitude in the $r_0\to\infty$ limit. The situation is different
in the case of higher dimensional $AdS$ space-times, {\it e.g.} for
$AdS_3$, where the boundary terms cannot be adjusted to
cancel all the divergences coming from the bulk terms,
reflecting
the effect of the central charge\cite{0506176}.}

Next we turn to the proof of $f(\theta)$ independence of the
QEF after inclusion of quantum corrections. 
For this we need to show that the
action evaluated for an off-shell field configuration is also
independent of the choice of $f(\theta)$. Now the allowed off-shell
field configurations over which we carry out the path integral are
constrained by the
boundary conditions on the various fields. We choose
the boundary conditions so as to allow only 
normalizable deformations
of the original background. In particular the difference
between
the action of the deformed
field configuration and the original action must be finite. This requires
that
the additional contribution $\Delta \left(\sqrt{\det g}\, \LL\right)$ 
in the 
lagrangian density,
that arises due to the difference between the general 
off-shell field
configuration and the original $AdS_2\times K$
background, must fall off faster
than $r^{-1}$ for large $r$. To see
if such a $\Delta \left(\sqrt{\det g}\, \LL\right)$ 
can give $f(\theta)$ dependent contribution
to the action we can compute the difference between the contributions
for a general $f(\theta)$ and $f(\theta)=1$. This is given by
\be \label{edd1}
\int_0^{2\pi} d\theta \int_{r_0}^{r_0 f(\theta)} \, dr\, 
\Delta \left(\sqrt{\det g}\, \LL\right)\, .
\ee
Since for large $r$, $\Delta \left(\sqrt{\det g}\, \LL\right)$ 
falls of faster than $r^{-1}$,
the integrand falls off faster than $r_0^{-1}$ in the domain of
integration. On the other hand the size of the domain is
of order $r_0$. Thus \refb{edd1}
vanishes in the $r_0\to\infty$ limit. This shows that even the
off-shell action is independent of the choice of $f(\theta)$.

Since the above argument has been somewhat abstract, we shall now
illustrate explicitly how it works for the terms involving the
metric, gauge fields and scalars.
We
choose the following boundary conditions on the metric and the
gauge fields (see {\it e.g.} \cite{0809.4264})
\ben \label{ew2}
&& g_{\eta\eta}=v + \OO(e^{-2\eta}), \quad 
g_{\eta\theta} =  \OO(e^{-2\eta}), \quad 
g_{\theta\theta}={v\over 4}\, e^{2\eta} +\OO(1), \nonumber
\\ &&
A^{(i)}_\theta = -{1\over 2}\,
i \, e_i\, e^\eta + \OO(1), \quad
A^{(i)}_\eta =  \OO(e^{-2\eta}), \qquad F^{(i)}_{\eta\theta}
= -{1\over 2}\,i\, e_i\, e^\eta +\OO(e^{-\eta})\, .
\een
This background differs from the $AdS_2$ background by terms
which fall off as powers of $e^{-\eta}$ for large $\eta$. We shall refer
to these terms as subleading terms.
We shall now
show that in the
$r_0\to\infty$ limit the contribution from the 
subleading terms in
\refb{ew2} to the action is independent of $f(\theta)$. 
On the other hand the leading contribution to the
action, being the same as that from an $AdS_2$ background, 
has already been shown to be independent of the choice of
$f(\theta)$. This would then establish that the off-shell action 
computed for an arbitrary configuration satisfying \refb{ew2} is
independent of $f(\theta)$ and 
hence the full QEF is also independent of $f(\theta)$. 

First we analyze the subleading contribution to $A_{bulk}$.
For this we express
the difference between the bulk action for a general 
$f(\theta)$ and $f(\theta)=1$ as
\be \label{ewx1}
\Delta A_{bulk}=
-\int_0^{2\pi} d\theta\, \int_{r_0}^{r_0 f(\theta)} dr \, \sqrt{\det g}\
\, \LL\, .
\ee
Now the integration volume is of order $r_0$ and the leading
order contribution to $\sqrt{\det g}\, \LL$ is of order unity. 
On the other hand the subleading terms 
in every field given in \refb{ew2} is
suppressed by a factor of $e^{-2\eta}$ compared to the leading
term. 
The only exception is the constant contribution
to $A^{(i)}_\theta$, but this does not contribute to the field
strength, and  $\LL$ depends on the $A_\mu^{(i)}$ only
through its field strength. Thus the net contribution to 
$\LL$ from the subleading
terms in \refb{ew2} is of order $e^{-2\eta}$ for large $\eta$.
This is of order $r_0^{-2}$ in the domain of integration given in
\refb{ewx1}, and
gives a 
contribution of order $r_0^{-1}$ to $\Delta A_{bulk}$.
This vanishes
in the $r_0\to\infty$ limit. 
There is of course a finite contribution to the
$A_{bulk}$ from the subleading terms from the region where $r$
(\i.e.\ $\eta$) is finite, but this does not depend on the choice
of $f(\theta)$.

An identical argument shows that the contribution from
the subleading terms to $A_{boundary}$ is suppressed by
powers of $r_0^{-1}$ both at $r=r_0$ and at $r=r_0\, f(\theta)$.
Finally the term proportional to $\int_0^{2\pi}
A^{(i)}_\theta\, d\theta$ may be analyzed as follows. We note that
\be \label{enote}
\int_0^{2\pi}A^{(i)}_\theta\, d\theta\bigg|_{r=r_0\, f(\theta)}
- \int_0^{2\pi}A^{(i)}_\theta\, d\theta\bigg|_{r=r_0}
= \int_0^{2\pi}\, d\theta\, \int_{r_0}^{r_0 f(\theta)}\, 
d\, r\, F^{(i)}_{r\theta}=\int_0^{2\pi}\, d\theta\, 
\int_{\cosh^{-1}r_0}^{\cosh^{-1}(r_0f(\theta))}\, d\eta\,
F^{(i)}_{\eta\theta}
\, .
\ee
Now since the subleading contribution to 
$F^{(i)}_{\eta\theta}$ is of order $e^{-\eta}\sim r_0^{-1}$ and the
$\eta$ integration spans a range of order unity, 
we conclude that the contribution to the right hand
side of \refb{enote} from the subleading terms in \refb{ew2}
is of order $r_0^{-1}$ and hence vanishes in the $r_0\to\infty$
limit.
This shows that in the $r_0\to\infty$ limit the contribution to the full
action from the subleading terms is independent of $f(\theta)$.

So far in our discussion we have ignored the scalar fields. If there are
massless scalar fields, then the natural boundary conditions on these
fields may be found by examining the 
solutions to the linearized  equations of motion
near the boundary and then allowing the normalizable 
mode to fluctuate. This gives, for a minimally coupled scalar field,
\be \label{escalar1}
\phi_s = u_s + \OO(e^{-\eta})\, ,
\ee
where $u_s$ is the attractor value. Thus
the subleading corrections to $\phi_s$ are suppressed by powers of
$e^{-\eta}$ instead of $e^{-2\eta}$, and could invalidate
our earlier argument. However since the attractor geometry extremizes
the action with respect to $\phi_s$, 
we expect that the correction to the action from the
subleading terms is at least
quadratic in $(\phi_s-u_s)$ and other
fluctuations and hence is again
suppressed by powers of $e^{-2\eta}$. 
Similarly the boundary terms in the action must also be quadratic
in the fluctuations and are suppressed by powers of $e^{-2\eta}$.
This in turn shows that the
off-shell action is independent of the choice of $f(\theta)$ even after the
inclusion of the subleading terms. Hence QEF is independent of the
choice of $f(\theta)$ even after inclusion of quantum corrections.

Finally we would like to note that our discussion has been centered
around studying the effect of integration over
the massless fields. In principle the contribution from integration
over the massive fields can be analyzed in the same way.
Alternatively we can  integrate out the
massive fields from the beginning and work with an effective
Lagrangian density $\LL_{eff}$ involving the massless fields.
Whatever we have done could then be repeated by replacing
$\LL$ by $\LL_{eff}$, and the final result -- that QEF is
insensitive to the details of the infrared cut-off -- would continue to
hold.

\sectiono{Effect of Asymptotic Symmetries} \label{easymp}

In this section we shall analyze the asymptotic symmetries of
string theory in the near horizon background of an extremal black
hole and their role in defining the path integral over the string
fields in the $AdS_2$
background\cite{0803.3621,0805.1861}. 
For this we consider the class of field configurations satisfying
\refb{ew2}, and 
identify diffeomorphisms accompanied by 
gauge transformations for which 
the asymptotic conditions given in \refb{ew2}
are preserved. The following diffeomorphism 
plus gauge transformation
satisfy this
restriction
\ben \label{ew3}
&& \theta = \chi(\wt\theta) - 2\, e^{-2\wt\eta} {\chi''(\wt\theta)} 
+\OO(e^{-4\wt\eta}), 
\qquad \eta = \wt\eta -\ln \chi'(\wt\theta) + \OO(e^{-2\wt\eta}),
\nonumber \\
&& \Lambda^{(i)} = -2\, i\, e_i \, e^{-\wt\eta} 
{\chi''(\wt\theta)\over \chi'(\wt\theta)} 
+\Lambda_0^{(i)}(\wt\theta)+ \OO(e^{-2\wt\eta})\, ,
\een
where 
$\chi(\wt\theta)$ and $\Lambda_0^{(i)}(\wt\theta)$ 
are some functions
satisfying
\be \label{ev1}
\chi(\wt\theta+2\pi)=\chi(\wt\theta)+2\, \pi\, , \qquad 
\Lambda_0^{(i)}(\wt\theta+2\pi)
=\Lambda_0^{(i)}(\wt\theta)\, , \qquad \chi'(\wt\theta)>0\, .
\ee
$\chi(\wt\theta)$ and $\Lambda_0^{(i)}(\wt\theta)$ generate global 
diffeomorphism
and global gauge transformations respectively. The subleading
terms in \refb{ew3} can be used to locally fix the gauge
\be \label{ew1}
g_{\wt\eta\wt\eta}=1, \quad g_{\wt\eta\wt\theta}=
A^{(i)}_{\wt\eta}=0\, ,
\ee
but we shall proceed without making any specific choice of gauge.
If we denote by $\wt \chi(\theta)$ the inverse transformation of $g$,
\i.e.\ $\wt\theta = \wt \chi(\theta)$, then we have
\be \label{ev6}
\wt \chi(\theta+2\, \pi) = \wt \chi(\theta) + 2\, \pi,
\qquad \wt \chi'(\theta) = 1/ \chi'(\wt\theta) > 0 \, ,
\ee
and the  transformations \refb{ew3} may be
inverted as
\ben \label{ev7}
&& \wt\theta = \wt \chi(\theta) - 2\, e^{-2\eta} 
{\wt \chi''(\theta)}+ \OO\left(e^{-4\eta}\right)\, ,
\qquad \wt\eta = \eta -\ln \wt \chi'(\theta) 
+ \OO(e^{-2\eta}), \nonumber \\
&& \wt\Lambda^{(i)} = - \Lambda^{(i)} =  - 2\, i\, e_i \,  e^{-\eta} 
{\wt \chi''(\theta)\over \wt \chi'(\theta) }
+\wt\Lambda_0^{(i)}(\theta)+ \OO(e^{-2\eta})\, ,
\een
where $\wt\Lambda_0^{(i)}(\theta)=-\Lambda_0^{(i)}(\wt\theta)$.

Let us now view \refb{ew3} as an active transformation and consider
two field configurations related to each other by such a
transformation. 
Naively the action, being diffeomorphism invariant,
will have the same value for these two configurations. 
However we should remember
that the action is divergent and must be regulated by putting a cutoff.
Thus for a fixed cutoff the new configuration 
generated by the transformation \refb{ew3}
has
{\it a priori} a different action than the one for the original background
given in \refb{et7}, \refb{et8}. We shall denote by $S[\chi]$ the
action associated with the new configuration with a cutoff at
$\wt\eta=\eta_0$, -- then the action associated with the original
background can be obtained by setting $\chi(\wt\theta)=\wt\theta$
in $S[\chi]$.
We can
compute $S[\chi]$ 
by using the transformation \refb{ev7} to map the configuration back
to the original configuration. This will change the cutoff to
\be \label{ev8}
\eta = \eta_0 + \ln \wt \chi'(\theta) + \OO(e^{-2\eta_0})\, ,
\ee
or equivalently, in the $r=\cosh\eta$ coordinate
\be \label{ev9}
r = r_0 \, \wt \chi'(\theta) + \OO(r_0^{-1}), \qquad 
r_0\equiv \cosh \eta_0\, .
\ee
Since the configuration expressed in the $(\eta,\theta)$ coordinate
system is $\chi$-independent, the only possible dependence of the action
on $\chi$ comes through the $\chi$ dependence of
the cutoff given in \refb{ev9}.
This is precisely the problem addressed in \S\ref{sinfra},
with $f(\theta)$ replaced by $\wt\chi'(\theta)$. 
In particular due to eq.\refb{ev6} we 
have $\int \, d\theta\wt\chi'(\theta)
=2\pi$, so that the last condition in \refb{ea1} is satisfied.
We can now use the result of \S\ref{sinfra} to infer that
$S[\chi]$ is independent 
of $\chi$.
This
allows us to declare diffeomorphisms of the type described
in \refb{ew3} as  gauge transformations
and restrict the path integral to over
configurations which are not related to each other by 
diffeomorphisms of the type given in \refb{ew3}.
In fact we are forced to do so, since otherwise summing over 
configurations related to each other by transformations \refb{ew3}
will produce an infinite factor in the path integral.
Similar remarks hold for gauge transformations
generated by $\Lambda^{(i)}(\theta)$.

We cannot however declare all transformations of the type 
\refb{ew3} as gauge transformations. If we do so then for
$AdS_2$ background, which is invariant under an
$SL(2,R)$ subgroup of the transformations \refb{ew3} generated
by
\be \label{esl2r}
\delta \, w = 
{i\over 2} 
(1+w)^2 \eps_{-1} -{1\over 2} 
(1-w^2) \eps_{0}-{i\over 2} 
(1-w)^2 \eps_1, \qquad
w\equiv \sqrt{r-1\over r+1}\, e^{i\theta}\, ,
\ee
we need to divide the path integral by the
volume of the $SL(2,R)$ group. Due to the infinite volume
of the $SL(2,R)$ group  the result will
vanish. To
remedy this we declare transformations of the type \refb{ew3},
modulo an $SL(2,R)$ transformation, as  pure gauge 
transformations.\footnote{In practice this can be achieved {\it e.g.}
by requiring the transformations \refb{ew3} to leave fixed three
points on the boundary of $AdS_2$.} In that case we do not need
to divide the path integral over $AdS_2$ background by the
volume of $SL(2,R)$ group, and this gives a finite result.

In supersymmetric theories there can be fermionic
symmetries which leave the action invariant,  and
we should
retrict the path integral so as not to integrate over these
symmetry directions; otherwise the path integral will vanish
due to integration over these fermionic zero modes.
However in this
case one could also try to use an alternate
approach in which the infinities arising from integration over the
bosonic zero modes may be canceled against the zeroes coming
from the integration over the fermionic zero modes. In this case we
do not need to declare the transformations \refb{ew3} and their
fermionic cousins as pure gauge transformations.

\sectiono{Index or Degeneracy?} \label{sindex}

One of the issues which arise in comparing the 
microscopic and the
macroscopic entropies is that in the microscopic theory we
typically compute the helicity trace
index while the Bekenstein-Hawking entropy
or Wald entropy is supposed to compute the logarithm of
the absolute degeneracy. 
So how can we compare the two quantities?
We shall now argue that quantum entropy
function formalism provides a natural resolution of this puzzle.
For definiteness we shall focus on four dimensional black holes,
but similar analysis can be carried out in other dimensions.
In four dimensions the relevant index is the helicity trace 
index\cite{9708062, 9708130} 
$B_{2k} = (-1)^k\, Tr\left[ (-1)^{2J} (2J)^{2k}\right] / (2k)!$ where
$J$ denotes the helicity of the state (or component of 
angular momentum along some specific direction in the rest frame)
and $4k$ is equal to the number of
supersymmetry generators which are broken by the black hole.
For quarter BPS dyons in $\NN=4$ supersymmetric string theories in
four dimensions, $4k=12$.

Now suppose that in \refb{e1} we replace the left hand side by
such a helicity trace index. Then on the right hand side also
we should compute the helicity trace index. Now the total
angular momentum $J$ carried by the black hole is a sum of the
angular momentum from the horizon and the hair. Thus the
$(-1)^{2J}(2J)^{2k}$ factor will be replaced by 
$(-1)^{2J_{hor}+2J_{hair}}(2J_{hor}+2J_{hair})^{2k}$.
The $(2J_{hor}+2J_{hair})^{2k}$ factor can be expanded in
binomial expansion and we get a sum of $2k+1$ different terms.
However only the $(2J_{hair})^{2k}$ term will
give a non-vanishing contribution to the trace. This is due to the
fact that typically the $4k$ fermion zero modes associated with the
$4k$ broken supersymmetry generators are all part of the hair
degrees of freedom\cite{0901.0359}. 
Saturating each pair of fermon zero modes
requires a factor of $(2J_{hair})$; thus 
we need $2k$ factors of $2J_{hair}$ to saturate
the fermion zero modes associated with the hair degrees of
freedom.
As a result on the right hand side of \refb{e1x} we now need to replace
$d_{hair}$ by the helicity trace $B_{2k;hair}$
involving the hair degrees
of freedom, and $d_{hor}$ by the Witten index
$Tr(-1)^{2J_{hor}}$ over the horizon degrees of freedom. 

Now QEF  computes the
degeneracy of states of a fixed angular momentum, just as
it  computes the degeneracy of states of 
a fixed charge, since from the point of view of the two dimensional
string theory living on $AdS_2$ the angular momentum can be
regarded as a component of the electric charge vector\cite{0606244}.
Let us denote by $d_{hor}(\vec q,J)$ the degeneracy computed
using QEF for fixed charges $\vec q$
and a fixed angular momentum
$J$ along the 3-direction. 
Then the contribution from angular momentum $J$ 
to the 
Witten index associated with the horizon is
$(-1)^{2J} d_{hor}(\vec q,J)$. 
Now  in four dimensions only
$J=0$ black holes are supersymmetric and contribute to the
index. Any other extremal black hole with $J\ne 0$ will break all
supersymmetries and hence will not contribute to the index.
This gives $d_{hor}(\vec q,J=0)$ as 
the contribution to the Witten index from the
horizon, and multiplying this by $B_{2k}$ associated with the hair
degrees of freedom we get the full contribution to the helicity
trace index. Thus eq.\refb{e1}, \refb{e1x} can be replaced by
\be \label{e1ind}
B_{2k;micro}(\vec q)=B_{2k;macro}(\vec q)\, ,
\ee
\be \label{e1indx}
B_{2k;macro}(\vec q) = \sum_n\,
\sum_{\{\vec q_i\}, \vec q_{hair}\atop \sum_{i=1}^n 
\vec q_i+ \vec q_{hair}=\vec q} 
\, \left\{\prod_{i=1}^n \, d_{hor}(\vec q_i, J_i=0)\right\}  \,
B_{2k;hair}(\vec q_{hair}; \{\vec q_i\})\, .
\ee
Since $d_{hor}(\vec q_i, J_i=0)$ is computed by QEF, 
eqs.\refb{e1ind}, \refb{e1indx} provide a way to compare the
helicity trace index in the microscopic description to the QEF in
the macroscopic description.

This point of view suggests that while comparing the indices on the
microscopic and the macroscopic sides we can not only compare their
magnitudes but also their signs. 
For example the sign of the helicity
trace index $-B_6$ for quarter BPS states
in a class of $\NN=4$ supersymmetric 
string theories was calculated in \cite{0708.1270} and was found to be
positive, at least 
in the limit when the charges are large. 
We can compare this with the macroscopic result for the index as
follows. For simplicity we shall consider the $\NN=4$
supersymmetric string theory obtained by
compactifying type IIB string theory
on $K3\times T^2$, and choose a black hole that carries only
D-brane charges (D5-branes wrapped on 5-cycles, D3-branes 
wrapped on 3-cycles and D1-branes wrapped on 1-cycles
of $K3\times T^2$) without any momentum,
fundamental string winding, KK monopole or H-monopole charges.
In this case we expect that the only hair degrees of freedom are 
the fermion zero modes associated with the 12
broken supersynmmetry generators, contributing a factor of unity
to $-B_{6;hair}$, since classical fluctuations of closed string fields
around the black hole background are not expected to produce any
D-brane charges. 
On the other hand
since $d_{hor}(\vec q,J=0)$ measures the ground state
degeneracy of the dual $CFT_1$ carrying quantum numbers
$(\vec q,J=0)$, it is a positive number. Thus $-B_{6;macro}$
computed from eq.\refb{e1indx} is positive, 
in agreement with the microscopic
result.

Another consequence of this viewpoint is that  the sum over
various configurations appearing on the right hand side of 
\refb{e1indx} involves only those configurations for
which the index associated with the hair is non-vanishing. In
particular any configuration with accidental fermion zero modes,
besides the ones associated with the broken supersymmetry
generators, will have vanishing index and hence will not contribute
to the sum. Such configurations include for example
multi-black hole solutions  in $\NN=4$ 
supersymmetric string theories
with at least one center describing a
large black hole. These are known not to contribute to the
index and hence are expected to have accidental fermion zero
modes.

This discussion also 
raises the question as to whether it is possible to directly
compare the microscopic and macroscopic degeneracies for a
given angular momentum instead
of the indices. While this should be possible in
principle, the lack of a non-renormalization theorem implies that
we need to carry out the microscopic computation directly
in a regime of the parameter space where gravity is strong enough
to produce a black hole. This is not possible with the currently
available techniques. Typically the microscopic spectrum is
computed in the weak coupling regime. By the time we turn on the
coupling and bring it to a region where the black hole description
is appropriate, the detailed information about the spectrum of BPS 
states in the microscopic theory may be lost except for an
appropriate index that is protected by supersymmetry.

\sectiono{Quantum Entropy Function for Torsion $>1$ Dyons}
\label{storsion}

In type IIB string theory compactified on $K3\times T^2$, 
or equivalently in heterotic string theory on $T^6$, the
charges carried by a generic dyon are labelled by a pair 
of 28 dimensional vectors $(Q,P)$, each belonging to the  
28 dimensional Narain lattice $\Lambda_{28}$
with signature (6,22)\cite{narain}.
Physically $Q$ and $P$ denote the electric and magnetic charge
vectors in the heterotic description.
It was shown in \cite{0801.0149} 
that with the help of S-duality 
transformations any charge vector can be brought to the form:
\be \label{ech1}
(Q,P)=(\ell \, Q_0, P_0), \quad \ell\in \ZZZ, \quad Q_0,P_0\in
\Lambda_{28}, \quad \gcd\{Q_{0i} P_{0j}
- P_{0i} Q_{0j}\}=1\, ,
\ee
where $Q_{0i}$, $P_{0i}$ are the components of
the vectors $Q_0$, $P_0$ along some primitive basis vectors of the
Narain lattice. 
The integer $\ell$ is a discrete duality invariant introduced in
\cite{0702150} which we shall refer to as torsion.

Since we shall be working in the type IIB description, it will be useful
to understand the interpretation of the charge vectors directly in
type IIB string theory. For this we represent $T^2$ as a product 
(topologically but not necessarily metrically) of
two circles $S^1\times \wt S^1$ and denote by $x^5$ and $x^4$
the coordinates along $S^1$ and $\wt S^1$ respectively, both 
normalized to have period $2\pi$. Furthermore
we shall restrict ourselves to configurations carrying only D-brane
charges, \i.e.\ D1/D3/D5-branes wrapped on $S^1$ or $\wt S^1$
times 0/2/4 cycles of $K3$. In this case the magnetic charge vector
$P$ measures winding numbers of various branes along $S^1$
and the 
electrtic charge vector
$Q$ measures winding numbers of various branes along $\wt S^1$.
In this limited subspace the charge vectors $Q$ and $P$
are 24 
dimensional instead of 28 dimensional, associated with the 24 even
cycles of $K3$. There is a natural metric in this 24 dimensional
space given by the intersection form of the even cycles of $K3$,
and this allows us to define inner products $Q^2$, $P^2$ and
$Q\cdot P$ among the charge vectors. We shall call these
continuous T-duality invariants.

Now one advantage of using pure D-brane configurations is that in
this case the hair modes are simple, -- they 
consist of just the twelve fermion zero modes associated with
broken supersymmetry generators, 
and do not carry any charge. This is due to the
fact that we do not expect classical fluctuations of closed string fields
to carry RR charges. Thus the hair modes give a contribution of
$1$ to $-B_{6,hair}$ and $d_{hor}(Q,P)$ can be equated to $-B_6$
of the full black hole. 
This can then be compared with the microscopic result for
$-B_6$.
The latter was computed in 
\cite{0802.0544,0802.1556,0803.2692} and takes
the form:
\be \label{ech2}
(-1)^{Q\cdot P+1}
\sum_{s|\ell} s\, f(Q^2/s^2, P^2, Q\cdot P/s)\, ,
\ee
where $(-1)^{Q\cdot P+1}
f(Q^2, P^2, Q\cdot P)$ denotes the $-B_6$ index of a dyon
of charge $(Q,P)$ with $\gcd\{Q_iP_j-Q_jP_i\}=1$. Since the
result for the index depends on the domain in which the
asymptotic values of the moduli field 
lie\cite{0702141,0702150,0706.2363,0806.2337}, \refb{ech2}
makes sense only if we specify the domain. 
We show in 
appendix \ref{sa} that \refb{ech2} holds if
for the $s$-th term we take the asymptotic moduli to coincide
with the attractor values for the charges $(Q/s, P)$. In this
case
$f$ appearing in \refb{ech2} can be regarded as the index associated
with single centered black holes\cite{0706.2363}.
The explicit form of $f(Q^2, P^2, Q\cdot P)$ involves a
Fourier transform of the inverse of the Siegel modular
form\cite{9607026,0412287,0505094,0506249,0605210}, and has
been given in appendix \ref{sa}.

Our goal in this
section will be to get an understanding of \refb{ech2} from the
quantum entropy function.
The function $f(Q^2,P^2, Q\cdot P)$ has the property that
for large $Q^2$, $P^2$, $Q\cdot P$ it behaves as
$\exp(\pi\sqrt{Q^2P^2-(Q\cdot P)^2})$. Thus the $s$-th term in
the sum will behave as $\exp(\pi\sqrt{Q^2P^2-(Q\cdot P)^2}/s)$.
The leading contribution in the large charge
limit, coming from the $s=1$ term, matches
the exponential of the Wald entropy
$\pi\sqrt{Q^2P^2-(Q\cdot P)^2}$, which, as we have seen,
can also be regarded as the classical contribution to
QEF.
This suggests that $s$-th term in the sum should arise from
the contribution to
QEF from a  new
saddle point --
satisfying the boundary condition described in
\refb{ew2} -- whose contribution to the finite part 
of the exponent in
\refb{ecc1} is $1/s$ times that
of the original $AdS_2\times S^2$ background. Our goal will
be find these saddle points. A non-trivial check will be that
these new saddle points should exist only for values of $s$ which
divide the integer $\ell$ defined in \refb{ech1}.

We begin by writing down the euclidean
near horizon metric associated with
this black hole. The requirement of $SL(2,R)$ symmetry, together
with the usual rotational and translational symmetries,
fixes the ten dimensional metric of type IIB string theory
to be of the form
\be \label{ech3}
ds^2 = v \, \left( {dr^2\over r^2-1}+(r^2-1)\,  d\theta^2\right)
+ w (d\psi^2 + \sin^2\psi d\phi^2) + {R^2\over\tau_2} 
\left|dx^4+\tau dx^5
\right|^2 + \wh g_{mn}(\vec u) du^m du^n\, , 
\ee
where $\wh g_{mn}$ and $\vec u$ denote the metric and the
coordinates along $K3$, $v$, $w$, $R$  are real constants
and $\tau=\tau_1+i\tau_2$ is a complex constant. 
$(r,\theta)$ label an Euclidean $AdS_2$ space whereas
$(\psi,\phi)$ label a 2-sphere.
Besides these the background 
contains fluxes of various RR fields. In the six dimensional
description, in which all the RR field strengths can be regarded
as self-dual or anti-self-dual 
3-forms after dimensional reduction on K3,
$Q$ represents
$RR$ fluxes through the 3-cycle spanned by $(x^5,\psi,\phi)$ and
$P$ represents
$RR$ fluxes through the 3-cycle spanned by $(x^4,\psi,\phi)$.
The (anti-)self-duality constraints on the various components of
the RR fields in six dimensions
relate the fluxes through the $(x^4,r,\theta)$ and
$(x^5,r,\theta)$ planes to those through the $(x^5,\psi,\phi)$
and $(x^4,\psi,\phi)$ planes.

Since $Q$ according to \refb{ech1} is $\ell $ 
times a primitive vector, the RR fluxes through the cycle 
spanned by $(x^5,\psi,\phi)$ is $\ell$ times a primitive flux.
Let us now consider an orbifold of the background
\refb{ech3} by the transformation
\be \label{ech4}
(\theta,\phi,x^5)\to \left(\theta+{2\pi\over s},\phi+{2\pi 
\over s},x^5
+{2\pi k\over s}
\right)\, , \quad k,s\in \ZZZ, \quad \gcd(s,k)=1\, .
\ee
Since the circle parametrized by 
$x^5$ is non-contractible, this is a
freely acting orbifold. At the origin $r=1$ of the $AdS_2$ space
we have a non-contractible 3-cycle spanned by $(x^5,\psi,\phi)$,
with the identification 
$(x^5,\psi,\phi)=(x^5+2\pi k/s, \psi, \phi+2\pi/s)$.
As a result of this identification the total flux of RR fields through
this cycle is $1/s$ times the original flux. Since the flux quantization
constraints require the fluxes through this new 3-cycle to be integers,
we see that this orbifold is an allowed configuration in string theory
only when $\ell/s$ is an integer.

We shall now show, following \cite{0810.3472}, 
that this configuration has the same asymptotic
behaviour as \refb{ech3} up to allowed correction terms of the type
described in \refb{ew2} and hence must be included as a new saddle
point in the
path integral that computes the QEF.
For this we take
$(s\theta, r/s)$ to be our new $(\theta,r)$ coordinates.
This generates the metric
\ben \label{ech5}
ds^2 &=& 
v \, \left( {dr^2\over r^2-s^{-2}}+(r^2-s^{-2})\,  d\theta^2\right)
+ w (d\psi^2 + \sin^2\psi d\phi^2) 
\nonumber \\ &&
+ {R^2\over \tau_2} \left|dx^4+\tau dx^5
\right|^2 + \wh g_{mn}(\vec u) du^m du^n\, , \nonumber \\
(\theta,\phi,x^5)&\equiv& \left(\theta+{2\pi},\phi+{2\pi\over s},x^5
+{2\pi k\over s}
\right)\, , \quad s|\ell\, .
\een
The RR field strengths in the new coordinate system remain identical
to those in the original coordinates.
Thus for large $r$ the metric, RR field strengths 
and the $\theta$ periodicity has the same
structure as \refb{ech3} but the $\phi$ and $x^5$ coordinates are
twisted by $2\pi/s$ and $2\pi k/s$ respectively
as we go around the $\theta$ circle. 
These can be regarded as the effect of switching on constant
Wilson lines at $\infty$ for the gauge fields associated with $\phi$
and $x^5$ translation symmetries\cite{0810.3472}, {\it without
changing the asymptotic values of the field strengths,
\i.e.\ charges.} Since in the path
integral we must integrate over 
the constant modes of the gauge fields keeping the charges
fixed,
\refb{ech5} represents an allowed configuration in the path 
integral {\it for the same values of the charges for which the
original solution \refb{ech3} is given}.

Since this new configuration is obtained by taking a $\ZZZ_s$
quotient of the original configuration, 
the action associated with this configuration
will be $1/s$ times the original action. But due to the rescaling
of the $r$ coordinate the cutoff $r_0$ on the new coordinate
$r$ will correspond to a cutoff $sr_0$ on the original radial
coordinate.
The net result is that the infrared divergent part of the action, being
proportional to $r_0$, is unchanged but the finite part gets scaled
by $1/s$. Thus in the large charge limit
the leading contribution to the QEF from this saddle point goes as 
$\exp(\pi\sqrt{Q^2P^2-(Q\cdot P)^2}/s)$. This observation,
together with the fact that these orbifolds exist only when $s$
divides $\ell$, makes this saddle point an ideal candidate that
contributes to the
$s$-th term in the sum in \refb{ech1}.

There is however a subtle issue related to this analysis.
The orbifold operation \refb{ech4} breaks the $SL(2,R)\times SU(2)$
isometry of $AdS_2\times S^2$ to a $U(1)\times U(1)$ subgroup.
Thus we actually have a
family of such orbifolds parametrized by the points on 
$(SL(2,R)/U(1))\times (SU(2)/U(1))$. Since $SL(2,R)/U(1)$
has infinite volume, it would suggest that the contribution from this
saddle point is multiplied by an infinite factor. However
we should note that the orbifold operation also breaks half of the 
eight supersymmetries of the original background, and hence has
four fermion zero modes. Integration over these fermion zero modes
produce zero result. This is precisely the problem analyzed in
\cite{0608021} who showed that by suitably regularizing the action one
gets a cancellation between the infinities coming from the bosonic
zero mode integrals and the zeroes coming from the fermion zero
mode integrals. Thus one gets
finite contribution from the saddle points described in \refb{ech5}.

By examining the symmetries of various terms
we can get further evidence for the identification of the $s$-th
term in \refb{ech2} with the contribution from the orbifold
\refb{ech4}. We first note that the function $f(Q^2,P^2,Q\cdot P)$
is invariant under an $SL(2,\ZZZ)$ transformation on the charges
\be \label{esl1}
\pmatrix{Q\cr P}\to \pmatrix{a & b\cr c & d} \, \pmatrix{Q\cr P}\, ,
\quad a,b,c,d\in \ZZZ, \quad ad-bc=1\, ,
\ee
which gives
\be \label{esq1}
\pmatrix{Q^2\cr P^2 \cr Q\cdot P} \to
\pmatrix{a^2 & b^2 & 2\, a\, b \cr c^2 & d^2 & 2\, c\, d\cr
a\, c & b\, d & a\, d + b\, c} \pmatrix{Q^2\cr P^2 \cr Q\cdot P} \, .
\ee
Thus the $s$-th term in \refb{ech2},
proportional to $f(Q^2/s^2, P^2, Q\cdot P/s)$,
will be invariant under the
transformation \refb{esq1} if we restrict $a,b,c,d$ to saisfy
\be \label{esl2}
a,c,d\in \ZZZ, \quad b\in s\ZZZ, \quad ad-bc=1\, .
\ee
This describes a $\Gamma^0(s)$ subgroup of $SL(2,\ZZZ)$.
We shall now show that this is precisely the symmetry of the
contribution from the orbifold \refb{ech4} {\it 
if we assume
that once a saddle point has been fixed, the 
contribution depends only on the continuous T-duality invariants
$Q^2$, $P^2$ and $Q\cdot P$ and not on the arithmetical properties
of the charges.}
To this end, note that
since $Q$ and $P$ represent winding charges of various
branes along the $x^4$ and $x^5$ directions respectively, the
$SL(2,\ZZZ)$ transformation acts on these coordinates as
\be \label{esl3}
\pmatrix{x^4\cr x^5} \to \pmatrix{a & b\cr c & d} \, 
\pmatrix{x^4\cr x^5}\, .
\ee
Thus acting on an orbifold given in \refb{ech4}, it takes it to another
orbifold with the same shift symmetries on the $(\theta,\phi)$ coordinates,
but a new set of shifts on the $(x^4,x^5)$ coordinates:
\be \label{esl4}
\pmatrix{x^4\cr x^5} \to \pmatrix{x^4\cr x^5} + 2\, \pi\,
\pmatrix{a & b\cr c & d} \, \pmatrix{0\cr k/s}
= \pmatrix{x^4\cr x^5} + 2\, \pi\, \pmatrix{bk/s \cr dk/s}\, .
\ee
Now with the restrictions described in \refb{esl2},
$b/s$ is an integer and hence the shift of $x^4$ in \refb{esl4}
is trivial. On the other hand we see from 
\refb{esl2} that $d$ and
$s$ must be coprime. Since $k$ and $s$ are also coprime, we see that
$d\, k$ and $s$ are coprime. Thus the shifts in $(x^4,x^5)$ given in
\refb{esl4} are of the same type as the one appearing in 
\refb{ech4},  with $k$ replaced by a new integer 
$k'=dk$ coprime
to $s$. Thus the sum of the contributions from all orbifolds of the
type given in \refb{ech4}, with $k$ running over different values,
should be invariant under the $\Gamma^0(s)$
transformation on $Q^2$, $P^2$ and $Q\cdot P$ described in
\refb{esq1}, \refb{esl2}. This establishes that both the
$s$-th term in \refb{ech2} and the contribution from the orbifold
\refb{ech4} are invariant under the same symmetry group
$\Gamma^0(s)$.

Finally we should add a word of caution.
While our analysis identifies a
specific class of extra contributions to QEF for
dyons of torsion $>1$, we have not shown that these are the only
additional contributions. There may be other effects which also
contribute to the difference between the degeneracy formul\ae\ for
the dyons of torsion 1 and dyons of torsion$>1$. As an example
we would like to mention $\ZZZ_s$ orbifolds of the type
discussed in \refb{ech4}, but with $k=0$, \i.e.\ without any shift
along the $x^5$ coordinate. These orbifolds have codimension
four fixed planes sitting at the
origin of $AdS_2$ and  the north or the south pole
of $S^2$. If we now consider the plane spanned by the $(x^5,\psi,\phi)$
coordinates and sitting at the origin of $AdS_2$, then the total RR flux
through this plane is given by $Q/s$ as discussed below \refb{ech4}.
However this does not require $Q$ to be quantized in units of $s$
since there may be additional RR flux sitting at the fixed points
at $\psi=0,\pi$
through which the above plane is required to pass. Thus these orbifolds
exist for all $Q$ and could be responsible for the subleading
contribution to the entropy of quarter BPS dyons even for $\ell=1$
states\cite{0810.3472}. 
Neverheless the details of the contribution from this
fixed point could depend on whether $Q/s$ is an integer or not since
for integer $Q/s$ one does not need to have any RR flux at the fixed
points. This then would be an additional source for the extra contributions
to the degeneracies of dyons of torsion $>1$ compared to the dyons
of torsion 1.

\bigskip

{\bf Acknowledgement:}
We would like to thank Nabamita Banerjee, Shamik Banerjee,
Borun Chowdhury, Atish Dabholkar,
Justin David,
Rajesh Gopakumar, Chethan Gowdigere, Rajesh
Gupta, Dileep
Jatkar, Ipsita Mandal,  Samir Mathur, 
Shiraz Minwalla and Yogesh Srivastava for useful
discussions. 

\appendix

\sectiono{The Dyon Degeneracy Formula for Type IIB String Theory 
on $K3\times T^2$} \label{sa}

According to \cite{0802.1556}
the index $-B_6$ of 
a torsion $\ell$ dyon  carrying charges of the form
$(Q=\ell\, Q_0, P)$ is given by 
\ben \label{e1.1}
d(Q,P) &=&
{(-1)^{Q\cdot P+1}} \sum_{s|\ell}  s^4\, 
\int_{i M_1-1/2}^{iM_1+1/2} d\crh
\int_{iM_2-1/(2s^2)}^{i M_2+1/(2s^2)} d\cs 
\int_{i M_3-1/(2s)}^{i M_3+1/(2s)} d\cv \, \nonumber \\ &&
e^{-i\pi (\cs Q^2 + \crh P^2 
+ 2 \cv Q\cdot P)} \, 
\Phi_{10}\left(\crh, s^2\cs 
, s\cv 
\right)^{-1}\, ,
\een
where $\Phi_{10}(\rho,\sigma,v)$ is the Igusa cusp form and 
\be \label{eatt}
M_1 = 2\, \Lambda\, {Q^2\over \sqrt{Q^2P^2 - (Q\cdot P)^2}}\, ,
\quad M_2 = 2\, \Lambda\, {P^2\over \sqrt{Q^2P^2 - (Q\cdot P)^2}}\, ,
\quad M_3 = -2\, \Lambda\, {Q\cdot P\over \sqrt{Q^2P^2 - 
(Q\cdot P)^2}}.
\ee
Here
$\Lambda$ is a sufficiently large positive real number. The
choice of integration contour given in \refb{eatt}
is valid if 
we choose the asymptotic moduli to be at 
the attractor point corresponding to the charges
$(Q,P)$.
For this choice of the integration contour we pick up the contribution
from single centered black holes only\cite{0706.2363}.

Now making a change of variables 
\be \label{echn}
\rho=\crh, \qquad \sigma=s^2 \cs, \qquad v = s \cv\, ,
\ee
we can express \refb{e1.1} as
\ben \label{erex}
d(Q,P) &=&
{(-1)^{Q\cdot P+1}} \sum_{s|\ell} s\, 
\int_{i M_1-1/2}^{iM_1+1/2} d\rho
\int_{is^2M_2-1/2}^{i s^2M_2+1/2} d\sigma 
\int_{i sM_3-1/2}^{i sM_3+1/2} dv \, \nonumber \\ &&
e^{-i\pi (\sigma Q^2/s^2 + \rho P^2 
+ 2 v Q\cdot P/s)} \, 
\Phi_{10}\left(\rho, \sigma 
, v 
\right)^{-1}\, .
\een
Since for torsion 1 dyons only the $s=1$ term in \refb{erex}
contributes we see that the $s$-th term has the structure of
the degeneracy of a torsion 1 dyon with continuous T-duality
invariants 
$(Q^2/s^2, P^2, Q\cdot P/s)$. The only issue is whether the
choice of integration contour agrees with that relevant for
the single centered black holes carrying the
invariants 
$(Q^2/s^2, P^2, Q\cdot P/s)$.  To this end we note that the
choice of contour for the $s$-th term in \refb{erex} may be
expressed as
\ben \label{enx}
&& M_1 = 2\, \wt\Lambda\, {(Q/s)^2\over 
\sqrt{(Q/s)^2P^2 - ((Q/s)\cdot P)^2}}\, ,
\quad s^2\, M_2 = 2\, \wt\Lambda\, {P^2\over 
\sqrt{(Q/s)^2P^2 - ((Q/s)\cdot P)^2}}\, , \nonumber \\
&& s\, M_3 = -2\, \wt\Lambda\, {(Q/s)\cdot P\over 
\sqrt{(Q/s)^2P^2 - ((Q/s)\cdot P)^2}}, 
\qquad \wt\Lambda \equiv s\, \Lambda\, .
\een
Comparing this with \refb{eatt} we see that this is indeed the
correct choice of contour for picking up contribution from
single centered black holes
carrying invariants
$(Q^2/s^2, P^2, Q\cdot P/s)$. Thus we can express \refb{erex}
as
\be \label{emm}
(-1)^{Q\cdot P+1}
\sum_{s|\ell} \, s\,  f(Q^2/s^2, P^2, Q\cdot P/s)\, ,
\ee
where $(-1)^{Q\cdot P+1}
f(Q^2,P^2,Q\cdot P)$ represents the index of a single
centered black hole of torsion 1 and continuous T-duality 
invariants
$(Q^2,P^2, Q\cdot P)$.

%\small
%\baselineskip 12pt
%\parskip -12pt

%\bigskip


\begin{thebibliography}{99}

%\bigskip

\bibitem{0805.0095}
  A.~Sen,
  ``Entropy Function and $AdS_2/CFT_1$ Correspondence,''
  arXiv:0805.0095v4 [hep-th].
  %%CITATION = ARXIV:0805.0095;%%

\bibitem{0809.3304}
  A.~Sen,
  ``Quantum Entropy Function from AdS(2)/CFT(1) Correspondence,''
  arXiv:0809.3304 [hep-th].
  %%CITATION = ARXIV:0809.3304;%%


\bibitem{9601029}
  A.~Strominger and C.~Vafa,
  ``Microscopic Origin of the Bekenstein-Hawking Entropy,''
  Phys.\ Lett.\ B {\bf 379}, 99 (1996)
  [arXiv:hep-th/9601029].
  %%CITATION = HEP-TH 9601029;%%  

\bibitem{0901.0359}
  N.~Banerjee, I.~Mandal and A.~Sen,
  ``Black Hole Hair Removal,''
  arXiv:0901.0359 [hep-th].
  %%CITATION = ARXIV:0901.0359;%%

\bibitem{narain}
  K.~S.~Narain,
``New Heterotic String Theories In Uncompactified 
Dimensions $<$ 10,''
  Phys.\ Lett.\  B {\bf 169}, 41 (1986).
  %%CITATION = PHLTA,B169,41;%%

\bibitem{0702150}
  A.~Dabholkar, D.~Gaiotto and S.~Nampuri,
  ``Comments on the spectrum of CHL dyons,''
  arXiv:hep-th/0702150.
  %%CITATION = HEP-TH/0702150;%%

\bibitem{0712.0043}
S.~Banerjee and A.~Sen, 
``Duality Orbits, Dyon Spectrum and Gauge Theory Limit of
Heterotic String Theory on $T^6$'',
arXiv:0712.0043 [hep-th].
%%CITATION = ARXIV:0712.0043;%%

\bibitem{0801.0149}
  S.~Banerjee and A.~Sen,
  ``S-duality Action on Discrete T-duality Invariants,''
  arXiv:0801.0149 [hep-th].
  %%CITATION = ARXIV:0801.0149;%%

\bibitem{0802.0544}
  S.~Banerjee, A.~Sen and Y.~K.~Srivastava,
  ``Generalities of Quarter BPS Dyon 
Partition Function and Dyons of Torsion
  Two,''
  arXiv:0802.0544 [hep-th].
  %%CITATION = ARXIV:0802.0544;%%

\bibitem{0802.1556}
  S.~Banerjee, A.~Sen and Y.~K.~Srivastava,
  ``Partition Functions of Torsion $>1$ Dyons in Heterotic
String Theory on $T^6$,''
  arXiv:0802.1556 [hep-th].
  %%CITATION = ARXIV:0802.1556;%%

\bibitem{0803.2692}
  A.~Dabholkar, J.~Gomes and S.~Murthy,
  ``Counting all dyons in N =4 string theory,''
  arXiv:0803.2692 [hep-th].
  %%CITATION = ARXIV:0803.2692;%%
  
\bibitem{9807056}
  G.~W.~Moore,
  ``Attractors and arithmetic,''
  arXiv:hep-th/9807056.
  %%CITATION = HEP-TH/9807056;%%

\bibitem{9807087}
  G.~W.~Moore,
  ``Arithmetic and attractors,''
  arXiv:hep-th/9807087.
  %%CITATION = HEP-TH/9807087;%%

\bibitem{0401049}
  G.~W.~Moore,
  ``Les Houches lectures on strings and arithmetic,''
  arXiv:hep-th/0401049.
  %%CITATION = HEP-TH/0401049;%%


\bibitem{9607026}
R.~Dijkgraaf, E.~P.~Verlinde and H.~L.~Verlinde,
``Counting dyons in N = 4 string theory,''
Nucl.\ Phys.\ B {\bf 484}, 543 (1997)
[arXiv:hep-th/9607026].
%%CITATION = HEP-TH 9607026;%%

\bibitem{0412287}
G.~L.~Cardoso, B.~de Wit, J.~Kappeli and T.~Mohaupt,
``Asymptotic degeneracy of dyonic N = 4 string states
and black hole
entropy,''
JHEP {\bf 0412}, 075 (2004) [arXiv:hep-th/0412287].
%%CITATION = HEP-TH 0412287;%%

\bibitem{0505094}
  D.~Shih, A.~Strominger and X.~Yin,
  ``Recounting dyons in N = 4 string theory,''
  JHEP {\bf 0610}, 087 (2006)
  [arXiv:hep-th/0505094].
  %%CITATION = JHEPA,0610,087;%%

\bibitem{0506249}
D.~Gaiotto,
``Re-recounting dyons in N = 4 string theory,''
arXiv:hep-th/0506249.
%%CITATION = HEP-TH 0506249;%%

\bibitem{0605210}
  J.~R.~David and A.~Sen,
  ``CHL dyons and statistical entropy function from D1-D5 system,''
  JHEP {\bf 0611}, 072 (2006)
  [arXiv:hep-th/0605210].
  %%CITATION = JHEPA,0611,072;%%

\bibitem{0810.3472}
  N.~Banerjee, D.~P.~Jatkar and A.~Sen,
  ``Asymptotic Expansion of the N=4 Dyon Degeneracy,''
  arXiv:0810.3472 [hep-th].
  %%CITATION = ARXIV:0810.3472;%%

\bibitem{9802109}
  S.~S.~Gubser, I.~R.~Klebanov and A.~M.~Polyakov,
  ``Gauge theory correlators from non-critical string theory,''
  Phys.\ Lett.\  B {\bf 428}, 105 (1998)
  [arXiv:hep-th/9802109].
  %%CITATION = PHLTA,B428,105;%% 
 
\bibitem{9802150}
  E.~Witten,
  ``Anti-de Sitter space and holography,''
  Adv.\ Theor.\ Math.\ Phys.\  {\bf 2}, 253 (1998)
  [arXiv:hep-th/9802150].
  %%CITATION = 00203,2,253;%%

\bibitem{9711200}
  J.~M.~Maldacena,
  ``The large N limit of superconformal field theories and supergravity,''
  Adv.\ Theor.\ Math.\ Phys.\  {\bf 2}, 231 (1998)
  [Int.\ J.\ Theor.\ Phys.\  {\bf 38}, 1113 (1999)]
  [arXiv:hep-th/9711200].
  %%CITATION = IJTPB,38,1113;%%

\bibitem{9809027}
  A.~Strominger,
  ``AdS(2) quantum gravity and string theory,''
  JHEP {\bf 9901}, 007 (1999)
  [arXiv:hep-th/9809027].
  %%CITATION = JHEPA,9901,007;%%

\bibitem{9708062}
  A.~Gregori, E.~Kiritsis, C.~Kounnas, N.~A.~Obers, 
  P.~M.~Petropoulos and B.~Pioline,
  ``R**2 corrections and non-perturbative 
  dualities of N = 4 string ground
  states,''
  Nucl.\ Phys.\ B {\bf 510}, 423 (1998)
  [arXiv:hep-th/9708062].
  %%CITATION = HEP-TH 9708062;%%

\bibitem{9708130}
  E.~Kiritsis,
  ``Introduction to non-perturbative string theory,''
  arXiv:hep-th/9708130.
  %%CITATION = HEP-TH/9708130;%%

 \bibitem{9307038}
  R.~M.~Wald,
  ``Black hole entropy in the Noether charge,''
  Phys.\ Rev.\ D {\bf 48}, 3427 (1993)
  [arXiv:gr-qc/9307038].
  %%CITATION = GR-QC 9307038;%%

\bibitem{9312023}
  T.~Jacobson, G.~Kang and R.~C.~Myers,
  ``On Black Hole Entropy,''
  Phys.\ Rev.\ D {\bf 49}, 6587 (1994)
  [arXiv:gr-qc/9312023].
  %%CITATION = GR-QC 9312023;%%

\bibitem{9403028}
  V.~Iyer and R.~M.~Wald,
  ``Some properties of Noether 
  charge and a proposal for dynamical black hole
  entropy,''
  Phys.\ Rev.\ D {\bf 50}, 846 (1994)
  [arXiv:gr-qc/9403028].
  %%CITATION = GR-QC 9403028;%%

\bibitem{9502009}
  T.~Jacobson, G.~Kang and R.~C.~Myers,
  ``Black hole entropy in higher curvature gravity,''
  arXiv:gr-qc/9502009.
  %%CITATION = GR-QC 9502009;%%

\bibitem{0506177}
  A.~Sen,
  ``Black hole entropy function and the attractor mechanism in higher
  derivative gravity,''
  JHEP {\bf 0509}, 038 (2005)
  [arXiv:hep-th/0506177].
  %%CITATION = HEP-TH 0506177;%%

\bibitem{0506176}
  P.~Kraus and F.~Larsen,
  ``Microscopic black hole entropy in theories with higher derivatives,''
  arXiv:hep-th/0506176.
  %%CITATION = HEP-TH 0506176;%%

\bibitem{0809.4264}
  A.~Castro, D.~Grumiller, F.~Larsen and R.~McNees,
  ``Holographic Description of $AdS_2$ Black Holes,''
  arXiv:0809.4264 [hep-th].
  %%CITATION = ARXIV:0809.4264;%%

\bibitem{0803.3621}
  T.~Hartman and A.~Strominger,
  ``Central Charge for $AdS_2$ Quantum Gravity,''
  arXiv:0803.3621 [hep-th].
  %%CITATION = ARXIV:0803.3621;%%

\bibitem{0805.1861}
  M.~Alishahiha and F.~Ardalan,
  ``Central Charge for 
  2D Gravity on AdS(2) and AdS(2)/CFT(1) Correspondence,''
  JHEP {\bf 0808}, 079 (2008)
  [arXiv:0805.1861 [hep-th]].
  %%CITATION = JHEPA,0808,079;%%

\bibitem{0606244}
  D.~Astefanesei, K.~Goldstein, R.~P.~Jena, A.~Sen and S.~P.~Trivedi,
  ``Rotating attractors,''
  JHEP {\bf 0610}, 058 (2006)
  [arXiv:hep-th/0606244].
  %%CITATION = HEP-TH 0606244;%%

\bibitem{0708.1270}
  A.~Sen,
  ``Black Hole Entropy Function, 
Attractors and Precision Counting of
  Microstates,''
  arXiv:0708.1270 [hep-th].
  %%CITATION = ARXIV:0708.1270;%%

\bibitem{0702141}
  A.~Sen,
  ``Walls of marginal stability and dyon spectrum in N = 4 supersymmetric
  string theories,''
  arXiv:hep-th/0702141.
  %%CITATION = HEP-TH/0702141;%%

\bibitem{0706.2363}
  M.~C.~N.~Cheng and E.~Verlinde,
  ``Dying Dyons Don't Count,''
  arXiv:0706.2363 [hep-th].
  %%CITATION = ARXIV:0706.2363;%%

\bibitem{0806.2337}
  M.~C.~N.~Cheng and E.~P.~Verlinde,
  ``Wall Crossing, Discrete Attractor Flow, and Borcherds Algebra,''
  arXiv:0806.2337 [hep-th].
  %%CITATION = ARXIV:0806.2337;%%

\bibitem{0608021}
  C.~Beasley, D.~Gaiotto, M.~Guica, L.~Huang, 
  A.~Strominger and X.~Yin,
  ``Why Z(BH) = |Z(top)|**2,''
  arXiv:hep-th/0608021.
  %%CITATION = HEP-TH/0608021;%%


\end{thebibliography}
\end{document}